\documentclass[lettersize,journal]{IEEEtran}
\usepackage{amsmath,amsfonts}
\usepackage{algorithmic}
\usepackage{algorithm}
\usepackage{array}
\usepackage[caption=false,font=normalsize,labelfont=sf,textfont=sf]{subfig}
\usepackage{textcomp}
\usepackage{stfloats}
\usepackage{url}
\usepackage{verbatim}
\usepackage{graphicx}
\usepackage{cite}
\begin{document}

\title{The Future of QKD Networks}

\author{Alin-Bogdan Popa and Pantelimon George Popescu ~\IEEEmembership{}
\thanks{Corresponding author: P.G.Popescu}
\thanks{A.B.Popa and P.G.Popescu are with National University of Science and Technology POLITEHNICA Bucharest (emails: \underline{alin\_bogdan.popa@upb.ro}, \underline{pgpopescu@upb.ro})}
\thanks{Manuscript received XXXXX; revised YYYYY.}}



\maketitle

\begin{abstract}

With the recent advancements in quantum technologies, the QKD market exploded. World players are scrambling to win the race towards global QKD networks, even before the rules and policies required by such large endeavors were even discussed. Several vendors are on the market, each with specific parameters and advantages (in terms of key rate, link range, KMS software, etc.), hence considerable effort is now made towards standardization. While quantum communications is expected to reach a market size of up to \$36B by 2040, the largest QKD initiative to date is EuroQCI, which, due to its sheer scale, is forcing the market to mature. Although building a QKD network is believed to be trivial today, inter-connecting federated networks on a global scale is a heavy challenge. We propose QKD virtual networks not only as a useful infrastructure abstraction for increased flexibility and granular security, but as an inevitable solution for several problems that future QKD networks will encounter on the way towards widespread adoption. 
\end{abstract}


\section{Introduction}
\IEEEPARstart{I}{n} the first 4 months of 2024 alone, there have been 9,478 publicly disclosed cybersecurity attacks, consisting of over 35.9 billion records breached, with the MOAB (Mother Of All Breaches - a compilation of 26 billion records from companies such as Tencent, X/Twitter, LinkedIn, Adobe, Dropbox, Telegram, and more) from January 2024 as a significant portion \cite{databreaches}. The next 10 years may bring even more: we are now witnessing an era of potential dramatic and impactful change \cite{forbesWhatsAhead}.

We are at the peak Moore's law. The original statement that the number of transistors (and consequently the computing power) per fixed cost silicon chip is doubling every 18 months has not held up in recent years. Moore himself (who sadly passed away in 2023) has made in 2015 the prediction that Moore's law will slow down and perhaps hit a ceiling in the following 10 years. Massive efforts have been done to find alternatives for sustaining the same technological progress that humanity has enjoyed for the last 100 years. Two of the most promising and heavily funded alternatives are parallel computing (i.e. large-scale GPU infrastructures - see the transformative success of software like ChatGPT, Gemini, or MidJourney) and quantum computing, a technology which leverages quantum phenomena (specifically, quantum entanglement) to provide computing speed ups. Quantum computing has been experimented practically since the 20th century, with the first realization of quantum entanglement in 1949 by Chien-Shiung Wu, the first experimental violation of the Bell inequality in 1972 by John Clauser, the first experimental quantum key distribution in 1992 by Charles Benett and Gilles Brassard, and the first integer factoring with Shor's algorithm in 2001 by L.M.K. Vandersypen in an IBM research group. In recent years, quantum computing has gained significant traction and massive funding, with roughly \$42 billion of public investments in quantum technology to date, and with up to \$2 trillion (yes, with a T!) total added economic value of quantum computing estimated by 2040 \cite{mckinsey2024quantum}.

In computational complexity theory, the relevant complexity class is $BQP$ (Bounded-error Quantum Polynomial-time) - the class of decision problems that a quantum computer solves in polynomial time with a fixed bounded error probability. It is known that $P \subseteq BQP$ (where $P$ is the class of decision problems that can be solved by a deterministic Turing machine in polynomial time) and that $BQP  \subseteq PSPACE$ (where PSPACE is the class of decision problems that can be solved by a Turing machine in polynomial space). The relationship between $BQP$ and $NP$ (where $NP$ is the class of decision problems solvable by a non-deterministic Turing machine in polynomial time - perhaps the most relevant complexity class to optimization or decision problems that humanity has to solve on a day-to-day basis) is yet unknown; however, it is strongly conjectured that $BQP$ contains problems that fall outside of $P$, for which quantum computing can provide significant (in some cases, exponential) speed up over their classical counterparts \cite{bqp}. Among other evidence of the quantum superiority, quantum algorithms exist that search unstructured data in $O(\sqrt{N})$ time (Grover's algorithm) and that efficiently solve (i.e. with an exponential speed up over the best known classical algorithms) the integer factoring and discrete logarithm problems (Shor's algorithm), the discrete Fourier transform (QFT algorithm), and solving linear equation systems (HHL algorithm).

\IEEEpubidadjcol

Shor's algorithm \cite{shor}, in particular, is expected to have massive impact on online communication, mainly because key exchange algorithms (the initial phase and building block of modern communication cryptography) rely on some variant of discrete logarithm (in the multiplicative group of integers modulo p in the case of the Classical Diffie-Hellman, and on elliptic curves over finite fields in the case of Elliptic Curve Diffie-Hellman). As such, once the hardware for quantum computers become powerful enough to run Shor's algorithm against useful encryption such as ED25519 or RSA2048 (which is currently expected to require roughly 400,000 qubits, quite a bit more than the state of art existing hardware which recently only surpassed the 1000-qubit milestone; however, recent research has shown that Shor's algorithm can be distributed on only 3,226 systems with 127 logical qubits \cite{distributedUnitary2022}, so we must nervously acknowledge that we're almost there), the security of all online communication would be put at severe risk. 

Considering the HNDL (Harvest Now, Decrypt Later) strategy, tremendous efforts have been done to move towards a quantum-safe alternative as soon as possible. While some of these efforts have gone on the path towards Post-Quantum Cryptography, the only solution known today which provides theoretical proof of unconditional security (that is, security which does not rely on assumptions of limited computation power of the adversary) is Quantum Key Distribution (QKD), which leverages quantum phenomena such as photon polarization measurement and no-cloning theorem (e.g. BB84, B92 protocols), quantum entanglement and entanglement monogamy (e.g. E92 protocol), or coherent light pulses (e.g. COW protocol). In the general case, a QKD link is a physical connection (usually through fiber optics or free-space optics) that runs a QKD protocol between two endpoints in order to produce a shared secret at the ends in an unconditionally-secure manner. The shared secret may then be used to establish a secure communication session (e.g. use it as seed for an encrypted VPN to video-call your quirky uncle)  or to encrypt data directly (e.g. send information-theoretically secure encrypted cat pictures that \textit{only} the intended recipient will ever be able to decrypt).

By connecting multiple such links via trusted nodes, one can create large-scale QKD networks, enabling use-cases that can go much wider \cite{qkdnetworksInternet}. Several governmental and scientific initiatives have been deploying QKD networks (e.g. DARPA QKD network in the US, SECOQC in Austria, SwissQuantum in Switzerland, Tokyo QKD Network in Japan, QUESS space mission for free-space optics QKD with the Micius satellite in China \cite{micius}). 

By far, the largest such initiative is the European Quantum Communication Infrastructure (EuroQCI) in Europe \cite{euroqci}, commenced in 2019 by the EU in partnership with all 27 EU member states the European Space Agency (ESA), with the purpose of building a pan-european QKD network that secures the communication of a large number of governmental and EU institutions (including locations in public administration, military, education, healthcare, data centers, research institutes, and more). Through the Digital Europe program and with the guidance of the PETRUS Consortium, a funding of 170M euros (plus another 170M from the national governments) has been granted for the first stage of EuroQCI, which involves each EU member state deploying a National QCI (NatQCI) on its premises. The upcoming Connecting Europe Facility (CEF) call \cite{cefCall}, which is scheduled to be published in Summer 2024 with the funding start in Summer 2025, will put an additional 90M euros towards interconnecting the NatQCIs via cross-border terrestrial QKD links, and via free-space links with the Eagle-1 satellite, scheduled to launch by February 2026 and developed under a partnership between ESA, the European Commission, and the Luxembourg-based satellite company SES.

However, with great power comes great responsibility, and with such large-scale endeavors come unique challenges that have to be addressed. In order to fully capitalize on the potential benefits of the inter-connected EuroQCI, we will have to solve problems such as automated QKD network self-discovery, automated configuration, hierarchical QKD node addressing, communication protocols, key scheduling and lifecycle management, and more - all while preserving the QKD property of utmost importance: unconditional security (an important milestone towards the Quantum Internet \cite{quantumInternet}). In this paper we address the challenge of virtual separation into sub-networks for fine-grained control over use-case key usage especially over inter-connected, federated networks (such as EuroQCI) by introducing the concept of QKD Virtual Networks (QVNets).

\section{QKD Virtual Links}

The necessity for logical links (as opposed to physical) arises from the existence of potential use-cases between nodes that are not directly connected. For example, in a case of a three-node network (A, B, C) with physical links between A-B and B-C, if a key forwarding mechanism exist such that A and C can also obtain unconditionally secure keys, we consider A-C a logical link.

\begin{figure}[H]
\includegraphics[width=0.5\textwidth]{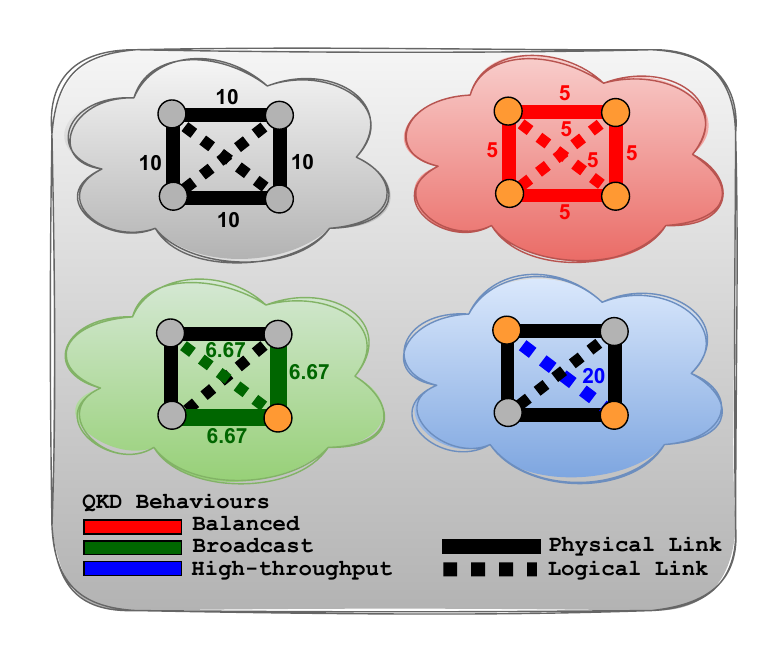}
\caption{A representation of a physical network (top-left), with the resulting key rate in three different scenarios: balanced key distribution (top-right), broadcast key distribution (bottom-left), high-throughput (bottom-right) - see \cite{balancedqkd}.}
\label{fig:pllink}
\end{figure}

A fundamental primitive is the concept of network behaviours as introduced by \cite{balancedqkd}, where the key forwarding and distribution may be configured to achieve different purposes depending on the needs of the QKD network administrators or stakeholders, as described in \ref{fig:pllink}. For example, in a federated and decentralized network, each node may require equitable access to the network keys, and as such the desired behaviour is a balanced one, where the minimum key rate between any pair of two nodes is maximized. In a centralized network, a broadcast scenario may be required, where one node (displayed in orange in \ref{fig:pllink}) needs a maximal key rate with all other nodes. Another possible behaviour is the one-to-one high-throughput, such as in a case of national emergency where the entire network needs to pause all key distribution in order to maximize the key rate between two given governmental institutions. Note that nodes connected physically are displayed in continuous lines, while logical links are displayed in dotted lines - either logical, physical, or both types of links may be part of the target links to be maximized during a given behaviour.

\begin{figure*}[t]
\centering
\includegraphics[width=\textwidth]{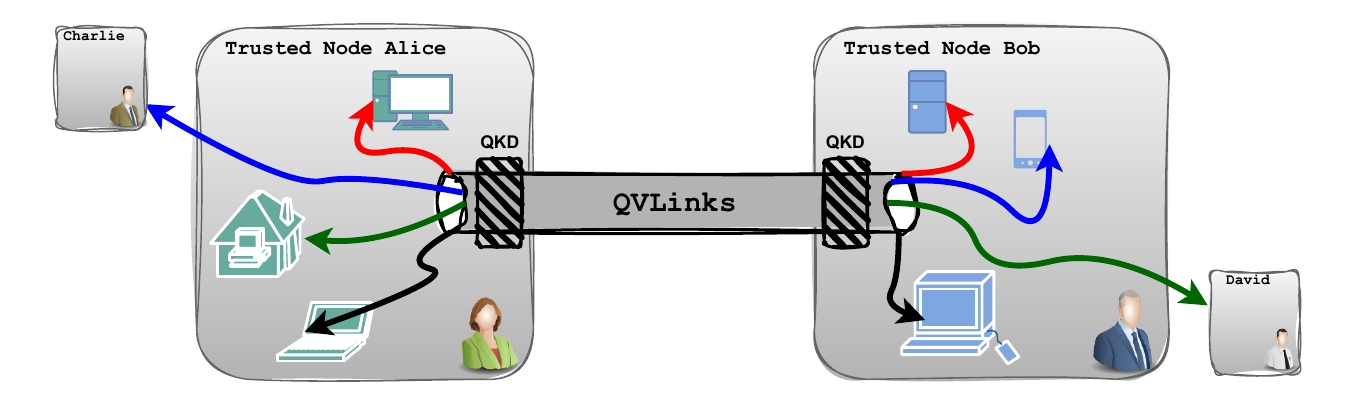}
\caption{A representation of a trunk logical link split into 4 QVLinks (red, blue, violet, black). QVLinks may be assigned different fractions of the total key rate available on the link (e.g. 1/2 for red, 1/4 for blue, 1/8 for violet, 1/8 for black). Each QVLink attends to a specific use-case, user, or sub-network. The number of QVLinks per logical link is unlimited, and their key rate quota can be dynamically adjusted based on the network conditions and key demand. Each QVLink may serve applications on the premises of Alice and Bob only, or may extend (perhaps through multi-hop physical links) towards other locations (as is the case for Charlie and David).}
\label{fig:trunk}
\end{figure*}

The fundamental unit of QVNets is the QKD Virtual Link (QVLink - see Fig. \ref{fig:trunk}). At the physical level, a typical QKD link consists of a physical channel which connects two QKD endpoints capable of running one QKD protocol in order to generate shared secrets. The shared secrets may be used on the spot, or may be aggregated in a key vault for later use. In practice, however, QKD hardware relies on photon-based communication, which, due to channel absorption and noise, has a limited range (for terrestrial links, typically around 60-120km \cite{qkdNetworking}); as such, connecting nodes over large distances may require several trusted repeaters (in the form of intermediary nodes) that forward the keys, typically via One Time Pad (OTP) by applying a XOR operation.

Moreover, this forwarding may be supported automatically by the Key Management System (KMS) software shipped by the vendor of the QKD hardware. As such, two nodes that are not directly connected in a QKD network, may still request and obtain shared keys with each other, with the forwarding and routing mechanisms under the hood hidden from the end-user. We consider such two nodes as being connected with a logical QKD link rather than a physical one.

The QVLink is the natural extension of the logical link, by considering each link as a trunk connection which can support multiple independent logical links. The motivation for this separation lies in the issue of limited key rates of commercially available QKD devices (which is typically expected to be around 1-4 kb/s; even though very few networks have achieved upwards of few hundred kb/s \cite{qkdNetworking}, the rate is still severely limiting the potential applications - and quantum funding would probably drop significantly if investors saw quantum-secure images loading slower than a dial-up connection in the '90s). As such, if several applications or use-cases or requesting personnel co-exist between the same two endpoints, then they necessarily compete for the limited resource that is the available key rate. By separating the key bandwidth into independent key streams, each stream can be assigned to different users or use-cases as necessitated by the network administrators. Additionally, programmatic rules can be put in place to adjust each stream's quota dynamically, depending on external conditions or key demand.

Formally, a logical link can be seen as a tuple $((v_A, v_B), r)$ where $(v_A, v_B)$ is an unordered pair of connected nodes, and $r$ is the key rate available on the link. We extend the logical link to the tuple $((v_A, v_B), r, k, C, f)$ where $k$ is the number of virtual sub-connections, $C$ is the set of sub-connection IDs with size $k$, and $f : C \rightarrow [0,1]$ is the sub-connection key rate function which assigns to each sub-connection by ID a fraction of the link's key rate $r$. Note that $f$ is not necessarily normalized: the key rates of the sub-connections may add up to less than $1$ (essentially using less than all of the available key rate on that link) or to more than $1$ (meaning that the quotas of the individual sub-connections exceed the available key rate on the link, and an additional resolution mechanism will need to be applied when the key requests clash with each other). A QVLink is then the tuple $((v_A, v_B), c, r_c)$, where $c \in C$ is the sub-connection ID, and $r_c = f(c) \times r$ is the sub-connection key rate quota.

\section{QKD Virtual Networks}

\begin{figure*}[t]
\centering
\includegraphics[width=\textwidth]{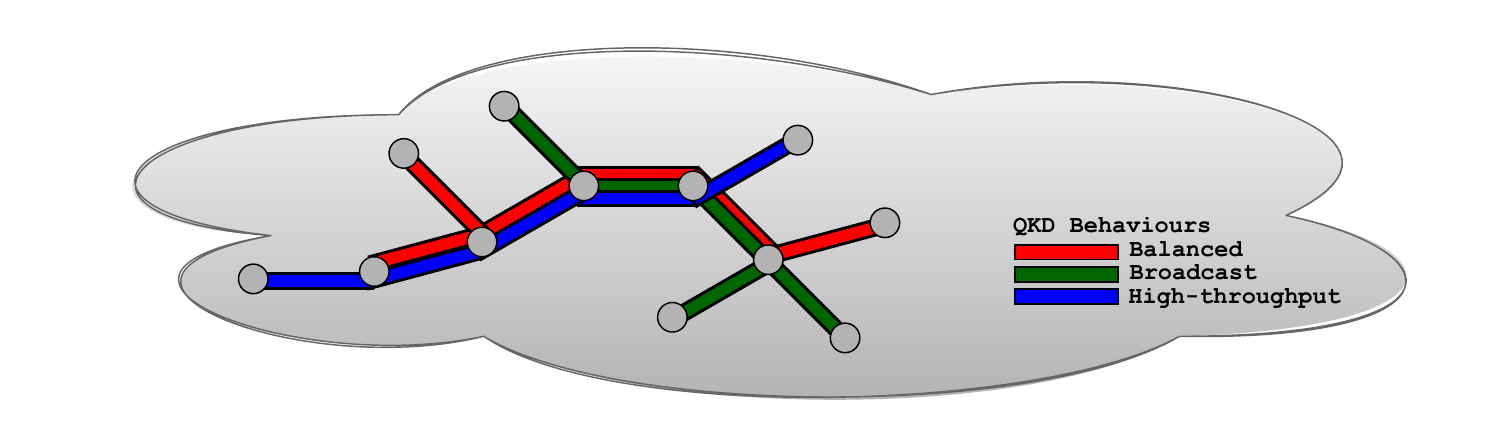}
\caption{A representation of QVNets over a physical QKD network. At the KMS level, different configurations may be imposed for each QVNet. For example, the KMS behaviours in terms of key forwarding for each QVNet may be balanced (forwading maximizes the minimum key rate between all pairs of nodes within that QVNet), broadcast (forwarding maximizes the minimum key rate between one fixed node and all others within that QVNet), and high-throughput (forwarding maximizes the key rate between two fixed nodes within that QVNet).}
\label{fig:qvnet}
\end{figure*}

With QVNets, we extend the QKD QVLink to the level of the network. A QVNet is the network graph composed of all QVLinks with the same ID (see Fig. \ref{fig:qvnet} for a representation). Formally, the QVNet is a subgraph of the original network graph, where the edge weight (i.e. key rate) is at most equal to the edge weight on the original graph. 

\begin{figure}[H]
\centering
\includegraphics[width=0.4\textwidth]{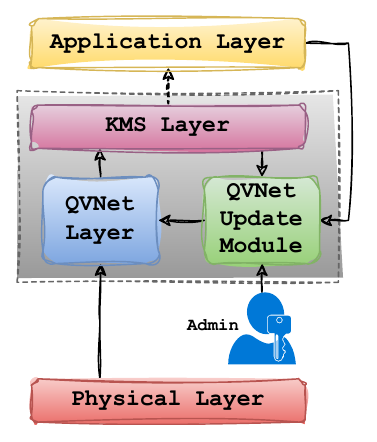}
\caption{Simple QKD architecture for QVNets}
\label{fig:layers}
\end{figure}

We view the network as composed of the following layers (see Fig. \ref{fig:layers}): 1) Physical layer (the actual physical devices and connections); 2) QVNet layer (further explained in this paragraph); 3) KMS layer (the KMS as designed by the network owners/administrators); 4) Application layer (the applications running on top of QKD which the end-users directly interact with). The KMS is designed and maintained by the network administrators, is perhaps made to be compatible with multiple vendors (especially for networks where all links are not acquired from the same vendor), and implements specific business rules as per the specific requirement of the organization(s) or government(s) that maintain and use the QKD network. The QVNet layer can be implemented, practically, as a policy layer right below the KMS layer, splitting the network graph into QVNet subgraphs as needed. The KMS would then run as multiple instances on each node, one for each QVNet the node is part of. Note that the separation into QVNets is purely virtual, and all the KMSs can be run physically on a single machine. It is important to note that QVNet rules can be updated dynamically based on demand for keys, applications running on top, and business rules. The update would be ordered by a QVNet Update Module, based on information collected from the KMS, applications, and the network administrators, hence creating a feedback loop from the KMS and Application layers back into the QVNet layer. Recent standardization work (see: ITU-T Y.3802, ETSI GS QKD 015, ETSI GS QKD 018) has focused on general KMS architecture and software-defined networks over QKD, but without support for virtual links and networks as described above. Additionally, since the QVNet layer is below the KMS, it is compatible with any software-defined \cite{commMagSDN} or virtual-key \cite{commMagVKey} strategy.

Each QVNet can be considered for the purposes of the KMS and any application running on top of the QKD infrastructure as a real physical network, analogue to VLANs in classical networks (we settled for \textit{QVNet} because \textit{VQAN} would sound rather odd). There are several immediate benefits: 1) Network separation and isolation. By imposing QVNets underneath the KMS layer, specific key rates on each link and nodes can be reserved for specific use-cases in order to prevent them from interfering with each other. 2) Flexibility and scalability. Massive cost efficiency can be obtained by reconfiguring QVNets purely on a software level depending on the needs of the network administrators and the desired use-cases rather than by changing the network on a hardware level, since commercially available QKD links cost in the range of several hundred thousand dollars. 3) Enhanced security and granular access. Since QKD links are typically maintained by large (public or private) organizations, there is a need for granular user access (either role-based or on a case-by-case basis). With QVNets, each user (or role, or group of users) may be given access to a specific QVNet and provided with a fixed maximum quota.

Again, an important issue is related to preserving unconditional security over a multi-hop physical connection. In a typical QKD network, each link produces shared keys at the ends; when a key is requested between two non-adjacent nodes, intermediary nodes need to use the intermediary keys (either by forwarding the key by OTP to the next node along the path, or by sending to a centralized KMS the result of the XOR operation between the keys with the previous and the next node). To preserve unconditional security, these keys cannot be reused later on; consequently, to produce one key between two non-adjacent nodes, a number of keys must be consumed that is equal to the number of physical QKD links along the chosen path between the two key-receiving nodes. Without proper access and key reservation rules, key consumption will be performed in a greedy manner on a first-come-first-served basis. A consequence of this is that when two nodes request a large number of keys, due to inherent limited key rates of QKD devices, they may inadvertently starve all intermediary nodes as well as all other pairs of nodes whose paths between each other pass through the two nodes. Think of an only elevator in a tall, pyramidal building that always prioritizes the lower floors (because more people live there) if multiple people call it at the same time: if you live on one of the upper floors, you either have to be prepared to wait, or renounce technology and take the stairs. QVNets are capable of mitigating this issue by setting strict quotas for groups of nodes/people as needed.

Moreover, depending on the requirements of the network owners, a specific behaviour may be needed on the level of the network. For example, recent research has shown that optimal key forwarding strategy can be achieved in a QKD network through linear programming for key reservation scenarios such as balanced (all-to-all), broadcast (one-to-all), or high-throughput (one-to-one) \cite{balancedqkd}. Other behaviours that may be needed to be imposed on a network include key routing rules (in the case of non-tree network graphs where multiple paths exist between the same two nodes), access rules (e.g. allowing a user or a group of users to request a specific number of keys or with specific nodes), or scheduling (e.g. enabling specific nodes or requests on some days only or during a specific time range). By logically splitting a network into QVNets, such behaviours or rules can be imposed separately on the level of each QVNet, providing more granular control and ease of configuration to the network administrators.

\begin{figure*}[t]
\centering
\includegraphics[width=\textwidth]{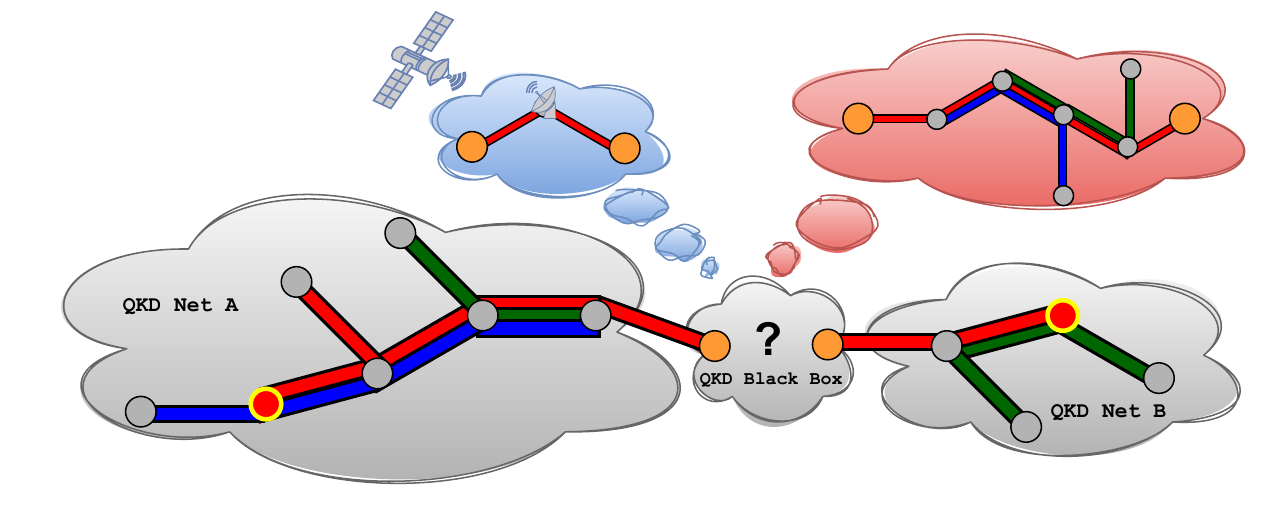}
\caption{Visualization of QVNet use-cases over a blackbox network. QKD networks A and B wish to exchange keys along the red path, but the path between them passes through the black-box network in the middle. Depending on the (potentially unknown) inner architecture of the black box, several situations may arise. Red: path between A and B requires terrestrial connection through the black-box network, which needs to co-exist with other networks within the black box. Blue: path between A and B requires satellite connection which is available via the black-box network on specific terms. The QVNets allow granular access for A and B through the black-box network. Not only this, but consider that you don't even know what the black box actually contains: the point of QVNets is that you don't need to.}
\label{fig:blackbox}
\end{figure*}

\section{QVNets over Blackbox Networks}

With the arrival of the CEF call for EuroQCI cross-border inter-connection, a common problem will be the inter-connecting of non-neighboring countries, since EuroQCI is a federated network in the sense that each NatQCI maintains a high level of autonomy over its network structure, choice of QKD vendors and devices, KMS software, forwarding/routing rules, etc. For example, the terrestrial path from country A to country B passes through country C; for both A and B, C is a blackbox (see Fig. \ref{fig:blackbox} for a visualization) of which they have no control (often times even the network structure and position of nodes are considered sensitive information of national security and are not exposed to the general public). There are cases where the connection from one country to the entire rest of mainland EU passes through specific countries without any alternative path (as is the case for Greece, whose terrestrial connection to the rest of EU necessarily passes through Bulgaria and Romania, as these are the only neighboring EU member states). In these cases, agreements can be negotiated between countries (on either pro bono or commercial terms) for intermediating key forwarding and ensuring a specific minimum key rate. This can be implemented practically by configuring a QVNet from A to B through C as intermediary, ensuring A and B can exchange keys as per the specific terms, and the rest of the network within C can use the remaining sub-network freely.

Part of the same upcoming CEF call, another issue arises due to prohibitive costs of deploying satellite connections. QKD-enabled satellites can be used for secure key exchange over large distances, and may even be the only viable approach for countries outside of mainland Europe (Ireland, Malta, Cyprus, etc.) where terrestrial connections may be unfeasible. Part of the second phase of EuroQCI, cross-country connections will be implemented with both terrestrial QKD links (via fiber optics) and with ground-satellite links via the prototype satellite Eagle-1 (with the long-term plan of deploying a larger mesh of QKD-enabled satellites for increased availability and key rate, most likely with the $IRIS^2$ Satellite Constellation). A satellite connection involves a) the satellite itself, whose orbit has to be scheduled depending on the position of the requesting node, as well as atmospheric and weather conditions, and b) the Optical Ground Station (OGS) which is capable of tracking the satellite, receiving the light pulses, decoding the key, and forwarding it to where it is needed within the terrestrial network. An OGS, however, while it appears to be little more than a fancy telescope with a light sensor (which it is), has a significantly large cost of deployment (in the range of 6M-7M euros for SES-built OGSs, although cheaper alternatives exist within the 1M-5M range). It is then expected that not all EU member states part of EuroQCI will be able to deploy their own OGS due to limited funding, and may have to rely on agreements with neighboring countries to share, borrow, co-buy, or co-build an OGS. Under these circumstances, deploying a QVNet between the partnering governments for granular access to the keys of the OGS would improve the usability of the space-segment of the network, by abstracting the implementation details and by integrating the business and political rules on a case by case basis within the QVNet protocol itself, below the KMS and Application layers. It is expected that in the long run, free-space and satellite communications will be indispensable for QKD networks, with proofs of concept already having been run in China \cite{micius}.

\section{Conclusion}

In the context of the threat of quantum computing, global players are striving to accelerate in the race towards quantum-safe communication. QKD networks provide a solution, but their current prohibitive costs and limited key rates require innovative solutions in order to enable them to support useful governmental and commercial use-cases. EuroQCI, the largest QKD initiative to date, is about to pose unique challenges with the arrival of the upcoming CEF call of 2024-2025, which consists of NatQCI inter-connecting through terrestrial cross-border QKD links and free-space links through the Eagle-1 satellite.

In this paper we propose a low-level protocol between the physical / Vendor KMS layers and the Network KMS, extending the concepts of logical links and VLANs from classical networks to the world of QKD, in the form of QVLinks and QVNets. We show how these can mitigate several issues of use-case clashing and cross-country blackbox routing, as well as increase the network's usability, flexibility, and cost efficiency.

For EuroQCI and particularly the soon-to-come cross-border connections, the problem of abstracting infrastructure for granular control (for which QVNets are a solution) is but one of the burning challenges that attract worldwide attention. Many other issues will need to be solved, such as node addressing, network discovery, automatic configuration, and more. We hope this is a needed step towards a global QKD network and the future quantum internet, paving the way for these technologies to be as ubiquitous and integral as the internet is today.

\section*{Acknowledgments}
This work is dedicated to professor Nicolae \c T\u apu\c s on his 75th anniversary.

This work has been partially supported by RoNaQCI, part of EuroQCI, DIGITAL-2021-QCI-01-DEPLOY-NATIONAL, 101091562.


\section{Biography Section}
 

\begin{IEEEbiography}[{\includegraphics[width=1in,height=1.25in,clip,keepaspectratio]{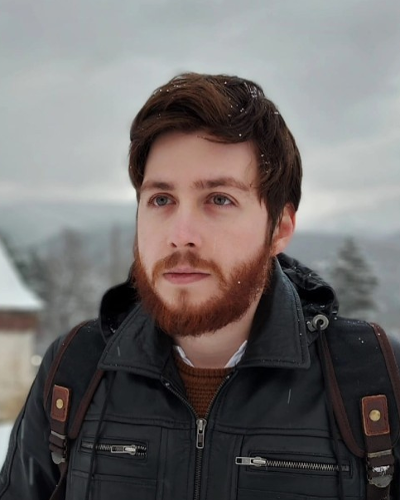}}]{Alin-Bogdan Popa} 
is a researcher at National University of Science and Technology POLITEHNICA Bucharest and a member of RoNaQCI (part of EuroQCI). He is CTO and partner at Parable, a venture studio and micro-to-small cap private equity fund with \$130M in assets-under-management. His main research interests are Quantum Communications, Decentralized Finance, Blockchain, Machine Learning. 
\end{IEEEbiography}

\begin{IEEEbiography}[{\includegraphics[width=1in,height=1.25in,clip,keepaspectratio]{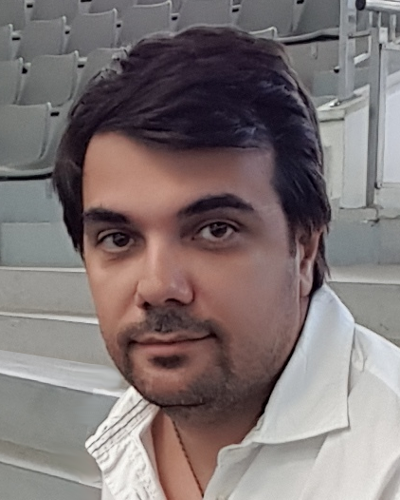}}]{Pantelimon George Popescu} is full Professor within the
Computer Science and Engineering Department and head of the Quantum Computing Laboratory at the National University of Science and Technology POLITEHNICA Bucharest. He is the Technical Coordinator of RoNaQCI (part of EuroQCI), member of the CERN QTI Advisory Board and member of the Open Quantum Institute Advisory Committee. His main fields of interest include Quantum Computing, Numerical Methods, Information Theory and Inequalities.
\end{IEEEbiography}

\vfill


\begin{thebibliography}{1}
\bibliographystyle{IEEEtran}

\bibitem{databreaches}
IT Governance, ``Global Data Breaches and Cyber Attacks in 2024'' [Online]. Available: https://www.itgovernance.co.uk/blog/global-data-breaches-and-cyber-attacks-in-2024. Accessed: June 10, 2024.

\bibitem{forbesWhatsAhead}
J. Prisco, "What’s ahead for quantum security in 2024?," Forbes Technology Council, Jan. 9, 2024. [Online]. Available: https://www.forbes.com/sites/forbestechcouncil/2024/01/09/whats-ahead-for-quantum-security-in-2024/. Accessed: Jun. 10, 2024.

\bibitem{mckinsey2024quantum}
McKinsey \& Company, ``Steady Progress in Approaching the Quantum Advantage,'' [Online]. Available: https://www.mckinsey.com/capabilities/mckinsey-digital/our-insights/steady-progress-in-approaching-the-quantum-advantage. Accessed: June 10, 2024.

\bibitem{bqp}
S. Aaronson, "BQP and the polynomial hierarchy," in \textit{Proceedings of the Forty-Second ACM Symposium on Theory of Computing}, Jun. 2010, pp. 141-150.

\bibitem{shor}
P. W. Shor, "Algorithms for quantum computation: discrete logarithms and factoring," in \textit{Proceedings of the 35th Annual Symposium on Foundations of Computer Science}, Santa Fe, NM, USA, Nov. 1994, pp. 124-134.

\bibitem{distributedUnitary2022}
A. Tănăsescu, D. Constantinescu, and P. G. Popescu, "Distribution of controlled unitary quantum gates towards factoring large numbers on today’s small-register devices," \textit{Scientific Reports}, vol. 12, no. 1, p. 21310, 2022.

\bibitem{qkdnetworksInternet}
Y. Cao, Y. Zhao, Q. Wang, J. Zhang, S. X. Ng, and L. Hanzo, "The evolution of quantum key distribution networks: On the road to the qinternet," \textit{IEEE Communications Surveys \& Tutorials}, vol. 24, no. 2, pp. 839-894, Apr. 2022.

\bibitem{micius}
S.-K. Liao, W.-Q. Cai, J. Handsteiner, B. Liu, J. Yin, L. Zhang, D. Rauch, M. Fink, J.-G. Ren, W.-Y. Liu, Y. Li, and J.-W. Pan, "Satellite-relayed intercontinental quantum network," \textit{Physical Review Letters}, vol. 120, no. 3, p. 030501, Jan. 2018.

\bibitem{euroqci}
European Commission, ``The European Quantum Communication Infrastructure (EuroQCI) Initiative'' [Online]. Available: https://digital-strategy.ec.europa.eu/en/policies/european-quantum-communication-infrastructure-euroqci. Accessed: June 10, 2024.

\bibitem{cefCall}
European Commission, ``Quantum communication infrastructure (EuroQCI)'' [Online]. Available: https://hadea.ec.europa.eu/programmes/connecting-europe-facility/about/quantum-communication-infrastructure-euroqci\_en. Accessed: June 10, 2024.

\bibitem{quantumInternet}
L. Gyongyosi and S. Imre, "Advances in the quantum internet," \textit{Communications of the ACM}, vol. 65, no. 8, pp. 52-63, Aug. 2022.

\bibitem{balancedqkd}
A.-B. Popa and P. G. Popescu, "Optimal Key Forwarding Strategy in QKD Behaviours," \textit{Scientific Reports} 14, 2024.

\bibitem{qkdNetworking}
M. Mehic, M. Niemiec, S. Rass, J. Ma, M. Peev, A. Aguado, V. Martin, S. Schauer, A. Poppe, C. Pacher, and M. Voznak, "Quantum key distribution: a networking perspective," \textit{ACM Computing Surveys (CSUR)}, vol. 53, no. 5, pp. 1-41, Sep. 2020.

\bibitem{commMagSDN}
A. Aguado, V. Lopez, D. Lopez, M. Peev, A. Poppe, A. Pastor, J. Folgueira, and V. Martin, "The engineering of software-defined quantum key distribution networks," \textit{IEEE Communications Magazine}, vol. 57, no. 7, pp. 20-26, Jul. 2019.

\bibitem{commMagVKey}
Y. Cao, Y. Zhao, J. Wang, X. Yu, Z. Ma, and J. Zhang, "KaaS: Key as a service over quantum key distribution integrated optical networks," \textit{IEEE Communications Magazine}, vol. 57, no. 5, pp. 152-159, May 2019.


\end{thebibliography}
\end{document}